\documentclass[
    ,final            ]
  {aipproc}

\layoutstyle{6x9}

\begin{document}

\title{Mueller-Navelet jets in high-energy hadron collisions}


\author{F.~Caporale}{
  address={Dipartimento di Fisica, Universit\`a della Calabria, and Istituto Nazionale di Fisica Nucleare, Gruppo collegato di
Cosenza, I-87036 Arcavacata di Rende, Cosenza, Italy}
}

\author{D.Yu.~Ivanov}{
   address={Sobolev Institute of Mathematics and
Novosibirsk State University, 630090 Novosibirsk, Russia}
}

\author{B.~Murdaca}{
  address={Dipartimento di Fisica, Universit\`a della Calabria, and Istituto Nazionale di Fisica Nucleare, Gruppo collegato di
Cosenza, I-87036 Arcavacata di Rende, Cosenza, Italy}
}

\author{A.~Papa}{
  address={Dipartimento di Fisica, Universit\`a della Calabria, and Istituto Nazionale di Fisica Nucleare, Gruppo collegato di
Cosenza, I-87036 Arcavacata di Rende, Cosenza, Italy}
}

\begin{abstract}

We consider within QCD collinear factorization the process $p+p\to {\rm jet}
+{\rm jet} +X$, where two forward high-$p_T$ jets are produced with a
large separation in rapidity $\Delta y$ (Mueller-Navelet jets~\cite{Mueller:1986ey}).
The hard part of the reaction receives large higher-order corrections $\sim \alpha^n_s (\Delta y)^n$, which can be
accounted for in the BFKL approach. We calculate cross section and azimuthal decorrelation, using the next-to-leading order jet vertices, in the small-cone approximation~\cite{small_cone}.

\end{abstract}

\maketitle

\section{Introduction}

Using collinear factorization, we study in the BFKL approach~\cite{BFKL} the production of two Mueller-Navelet jets in proton-proton collision $\big($proton($p_1$) + proton($p_2$) $\to$ jet($k_{J_1}$) + jet($k_{J_2}$) + $X$$\big)$ in the kinematical region where the jets are separated by a large interval of rapidity, $\Delta y \gg 1$, which means large center of mass energy $\sqrt s$ of the proton collisions. 

The results of a complete NLA analysis of the process under consideration were reported in~\cite{Colferai}, see also~\cite{DSW}. This numerical study followed previous ones~\cite{Sabio_Vera} based on the inclusion of NLA effects only in the Green's functions.
In our work the observables under study, the cross section and the moments of the azimuthal decorrelation, have been calculated in the full NLA BFKL approach, taking the convolution of the BFKL Green's function with the jet vertices calculated in the ``small-cone'' approximation~\cite{small_cone}.
\\
In the following we will show the dependence of our observables on the rapidity separation $\Delta y \equiv Y=\ln\frac{x_{J_1} x_{J_2} s}{|\vec k_{J_1}||\vec k_{J_2}|}$, where $s=2\left( p_1\cdot p_2\right)$, while $x_{J_{1,2}}$ and $\vec k_{J_{1,2}}$ are the longitudinal fractions and the jet transverse momenta, respectively.
\\
In QCD collinear factorization the cross section of the process reads
\vspace{-0.2cm}
\[
\frac{d\sigma}{dx_{J_1}dx_{J_2}d^2\vec k_{J_1}d^2\vec k_{J_2}}
=\sum_{i,j=q,\bar q,g}\int\limits^1_0 dx_1 \int\limits^1_0 dx_2\, f_i(x_1,\mu_F)
f_j(x_2,\mu_F) \frac{d\hat \sigma_{ij}(x_1 x_2 s,\mu_F)}{dx_{J_1}dx_{J_2}d^2
\vec k_{J_1}d^2 \vec k_{J_2}}\;,
\]
where the $i,j$ indices specify parton types (quarks $q=u,d,s,c,b$; antiquarks
$\bar q=\bar u,\bar d,\bar s,\bar c,\bar b$; or gluon $g$), $f_i(x,\mu_F)$
denotes the initial proton parton density function (PDF), the longitudinal
fractions of the partons involved in the hard subprocess are $x_{1,2}$,
$\mu_F$ is the factorization scale, $d\hat \sigma_{ij}$ is the
partonic cross section for the production of jets and $\hat s=x_1 x_2 s$ is
the squared center-of-mass energy of the parton-parton collision.
\\
The moments of the azimuthal decorrelations are given by
\[
\langle \cos[m(\phi_{J_1}-\phi_{J_2}-\pi)] \rangle \equiv
\langle \cos(m \varphi) \rangle = {\cal C}_m/{\cal C}_0\;,
\]
\[
{\cal C}_m = \int_0^{2\pi}d\phi_{J_1}\int_0^{2\pi}d\phi_{J_2}\,
\cos[m(\phi_{J_1}-\phi_{J_2}-\pi)] \,
\frac{d\sigma}{dy_{J_1}dy_{J_2}\, d|\vec k_{J_1}| \, d|\vec k_{J_2}|
d\phi_{J_1} d\phi_{J_2}}\;,
\]
where $ y_{J_i}=\frac{1}{2}\ln(x^2_{J_i} s/\vec k_{J_i}^2) $, $i=1,2$, are the jet rapidities in the center of mass system.
\begin{figure}[tb]
\centering
\includegraphics[scale=0.43]{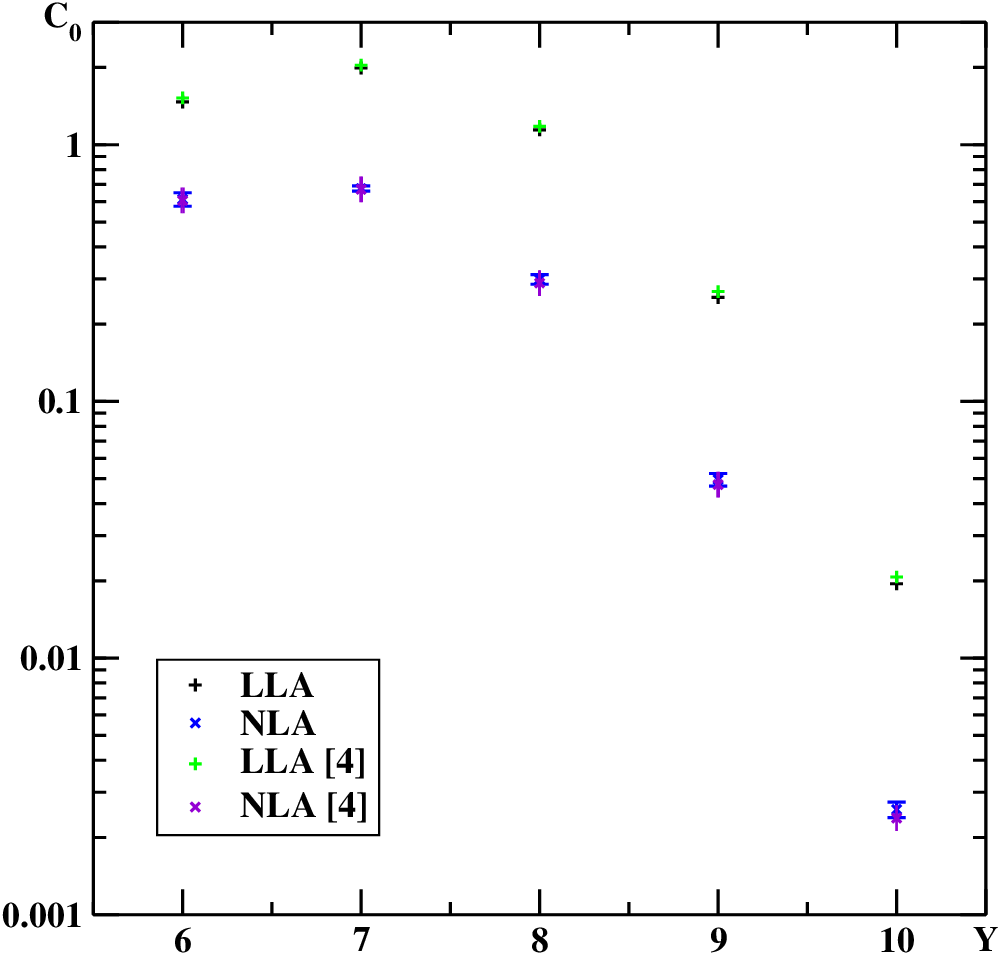}
\hspace{0.48cm}
\includegraphics[scale=0.43]{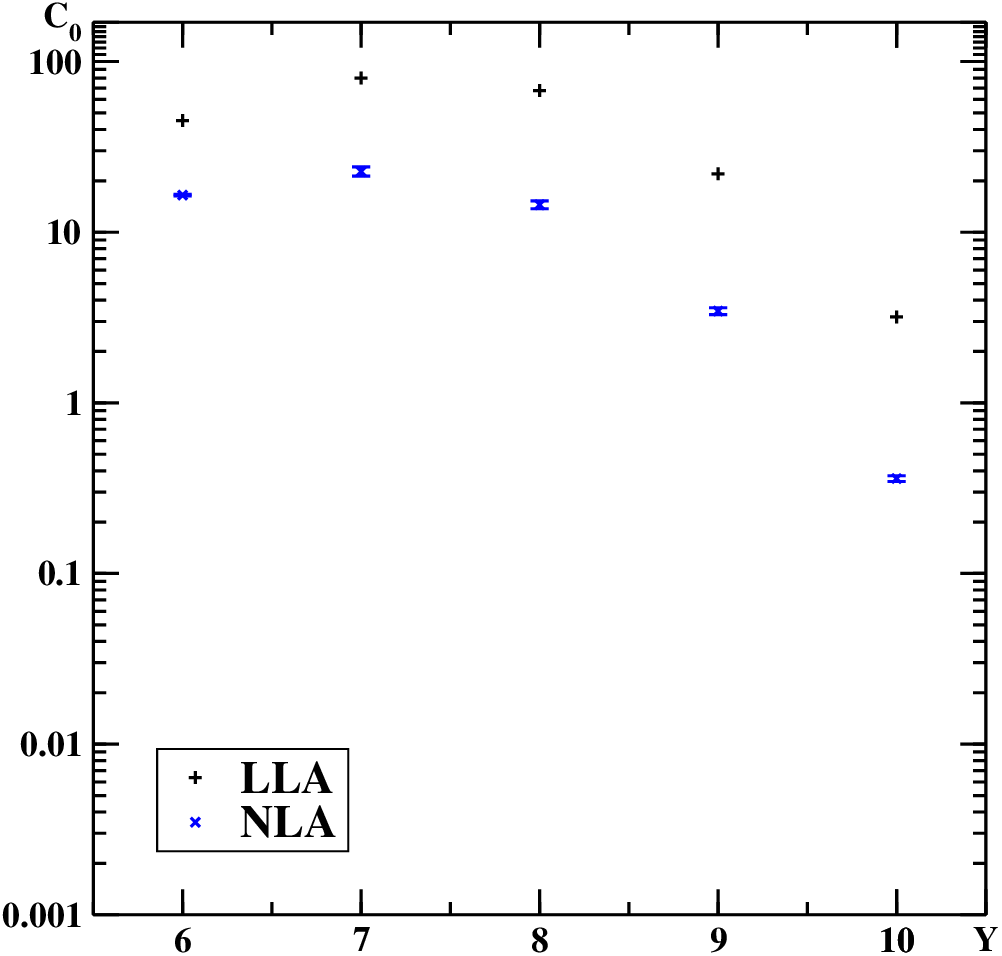}
  \caption{$Y$-dependence of the cross section ${\cal C}_0$
for $|\vec k_{J_{1,2}}|=35$ GeV (left) and $|\vec k_{J_{1,2}}|=20$ GeV (right). The typical optimal value for $\mu_R$ is around $3{\sqrt{|\vec k_{J_1}| |\vec k_{J_2}|}}$, while values of $Y_0$ are in the range 1--4.}
\vspace{-0.5cm}
\label{fig:C035}
\end{figure}
\\
In the NLA BFKL approach~\cite{BFKL}, the cross section of the hard process reads
\[
\frac{d\hat \sigma_{ij}(x_1 x_2 s,\mu)}{dx_{J_1}dx_{J_2}d^2
\vec k_{J_1}d^2 \vec k_{J_2}}=\frac{s}{(2\pi)^2}\int\frac{d^2\vec q_1}{\vec
q_1^{\,\, 2}} V_i(\vec q_1,s_0,x_1;\vec k_{J_1},x_{J_1})
\int\frac{d^2\vec q_2}{\vec q_2^{\,\,2}}V_j(-\vec q_2,s_0,x_2;
\vec k_{J_2},x_{J_2})
\]
\vspace{-0.3cm}
\begin{equation}
\times \int\limits^{\delta +i\infty}_{\delta
-i\infty}\frac{d\omega}{2\pi i}\left(\frac{x_1 x_2 s}{s_0}\right)^\omega
G_\omega (\vec q_1, \vec q_2)\, .
\label{hard}
\end{equation}
Here $V_{i,j}$ is the jet vertices
(impact factors) which we take in the small-cone approximation~\cite{small_cone}. The Green's function in~(\ref{hard}) obeys the BFKL equation~\cite{BFKL}. The artificial scale $s_0$ is introduced in the BFKL approach 
to perform the Mellin transform from the $s$-space to the complex angular
momentum plane and cancels in the full expression for the amplitude with the NLA accuracy.
\begin{figure}[tb]
\centering
\includegraphics[scale=0.43]{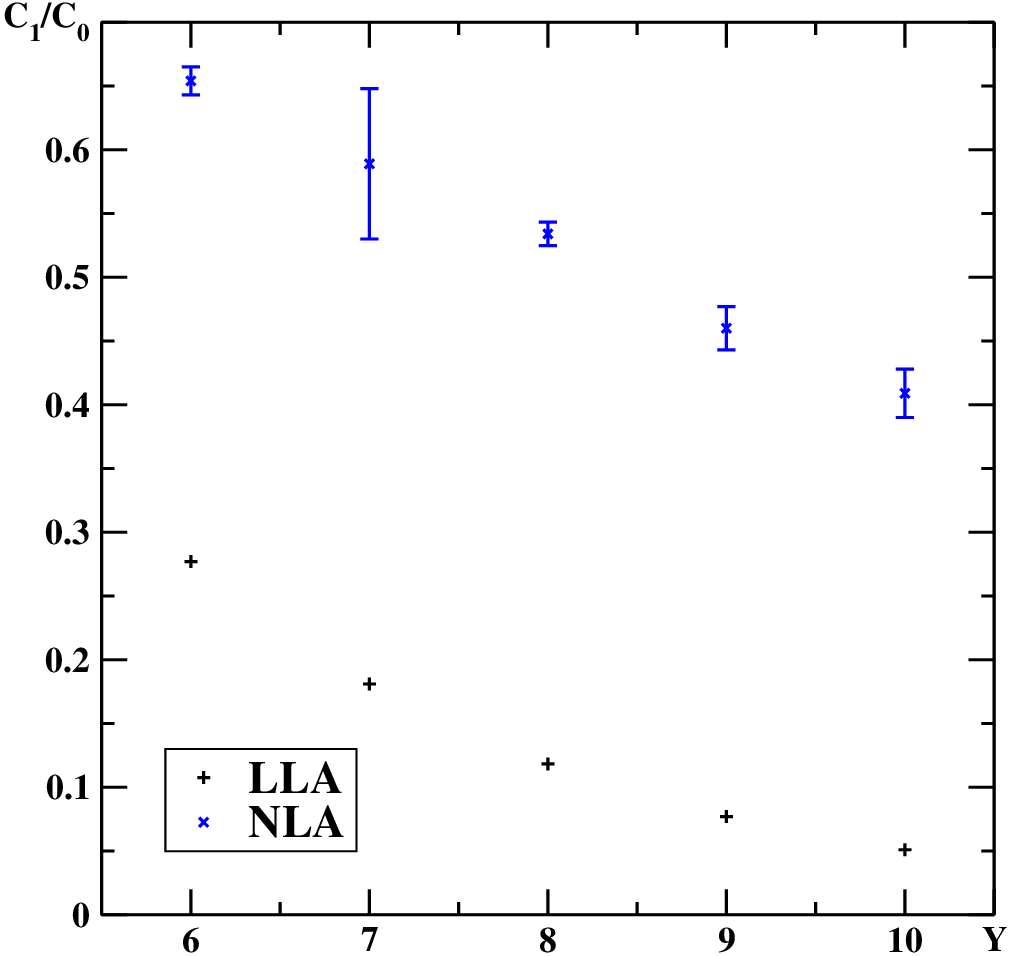}
\hspace{0.42cm}
\includegraphics[scale=0.43]{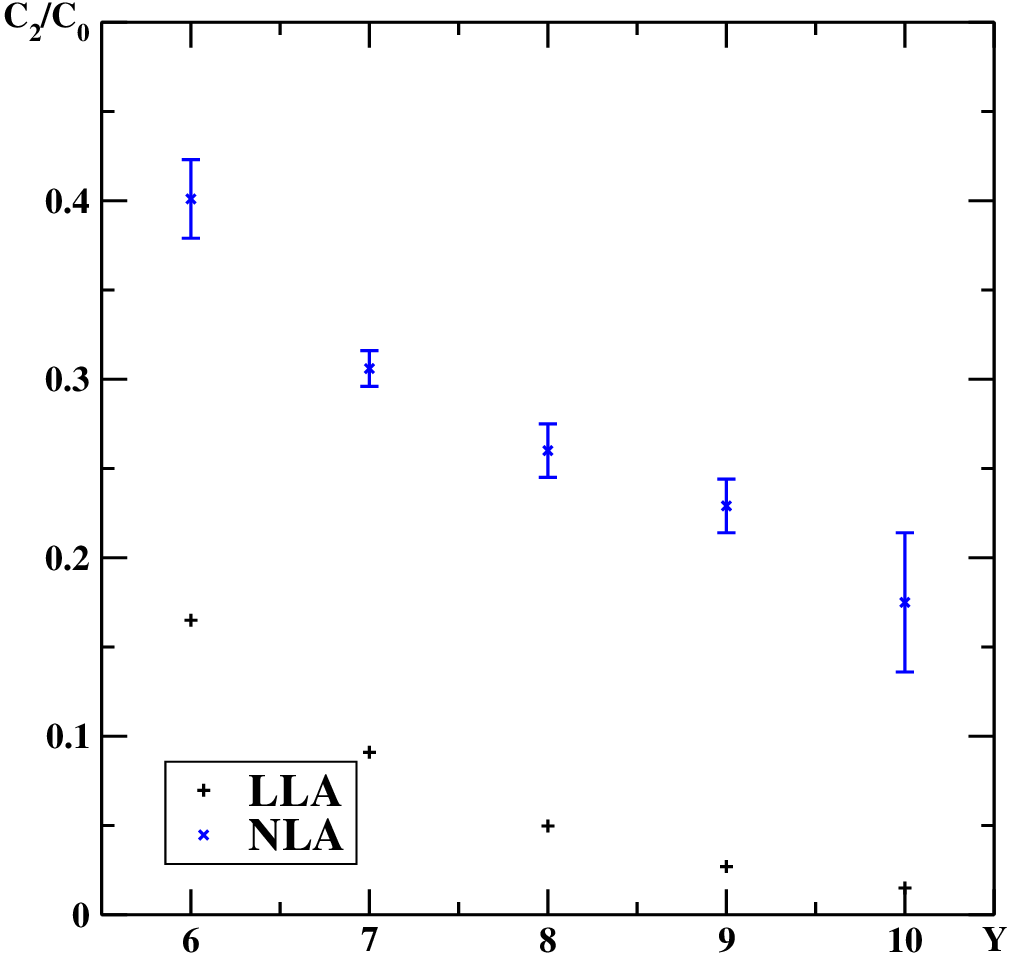}
\caption{$Y$-dependence of ${\cal C}_1/{\cal C}_0$ (left)
and ${\cal C}_2/{\cal C}_0$ (right) for $|\vec k_{J_1}|=|\vec k_{J_2}|=20$ GeV.}
\end{figure}
\vspace{-0.2cm}

\section{Numerical results}
\begin{figure}[tb]
\centering
\includegraphics[scale=0.43]{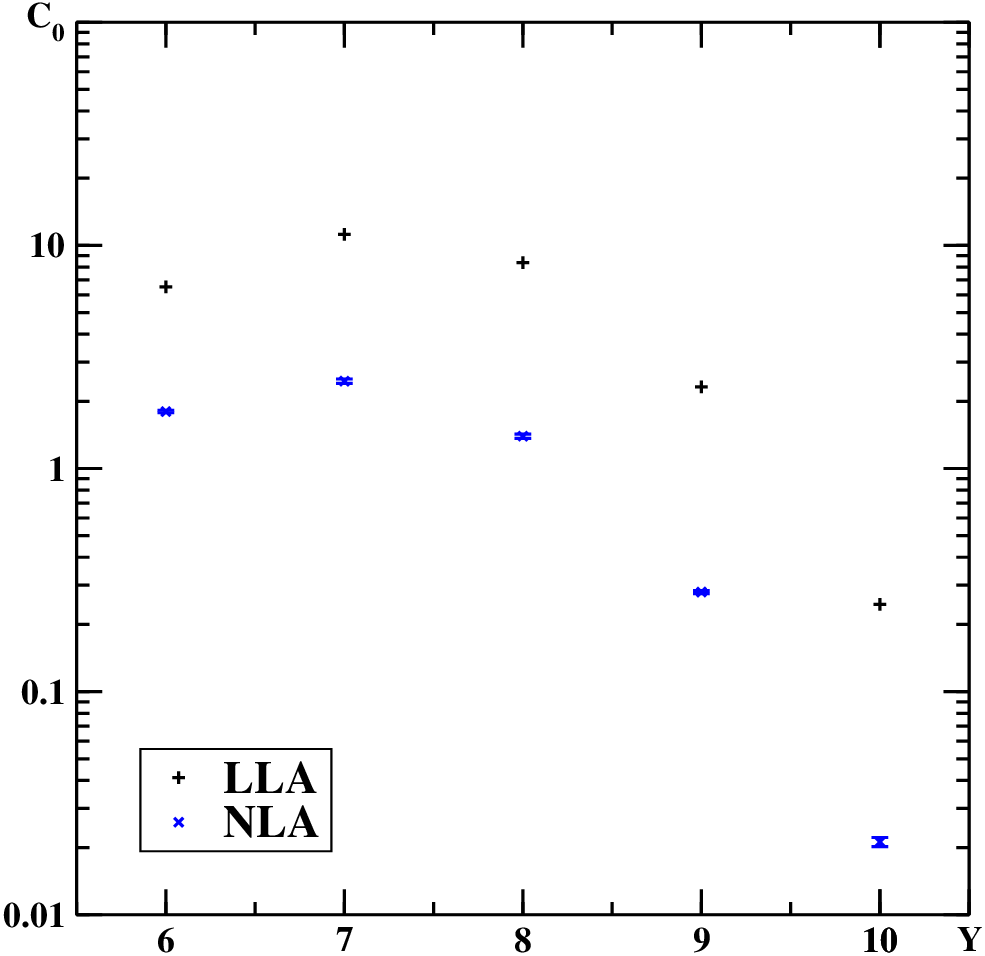}
\hspace{0.42cm}
\includegraphics[scale=0.43]{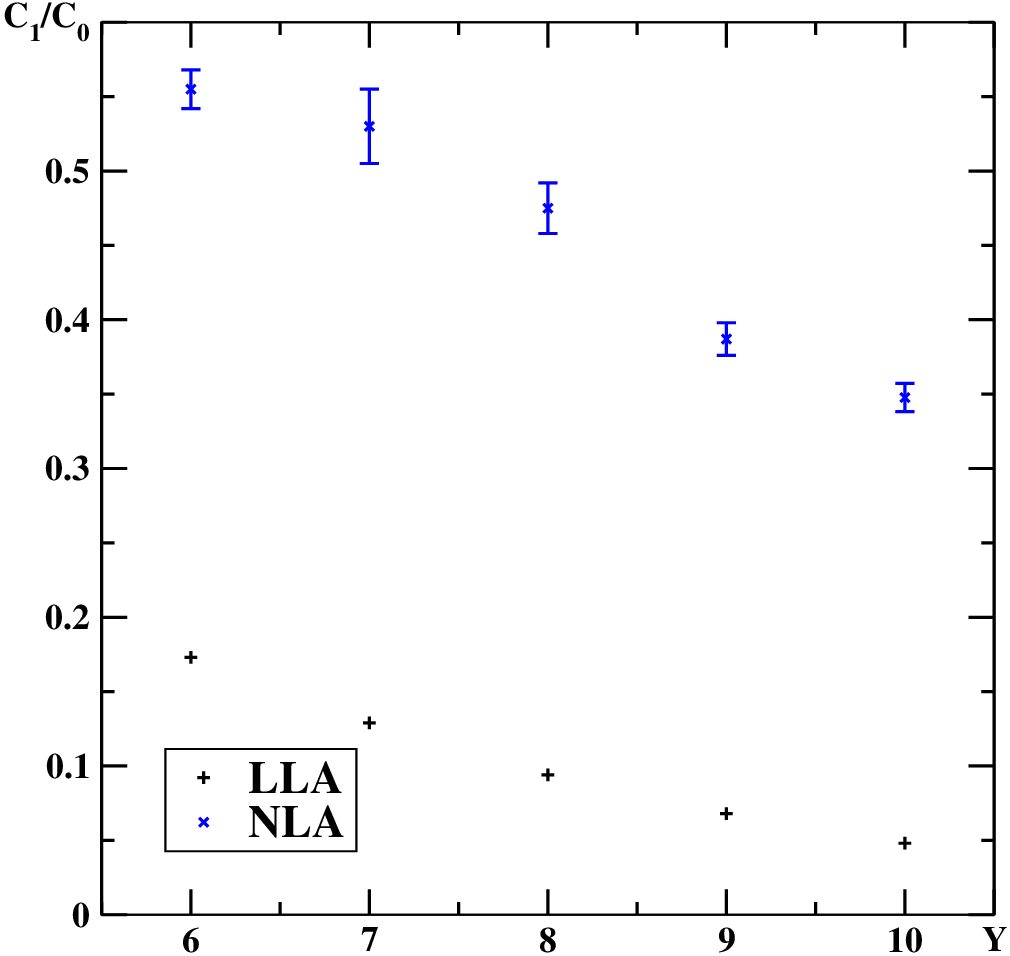}
\caption{$Y$-dependence of the  cross section ${\cal C}_0$ (left) and ${\cal C}_1/{\cal C}_0$ (right)
for $|\vec k_{J_1}|=20$ GeV and $|\vec k_{J_2}|=35$ GeV.}
\label{fig:C02035}
\end{figure}

We present our results for the dependence on $Y=y_{J_1}  - y_{J_2}$ of the 
${\cal C}_m$ calculated using exponentiated representation~\cite{exponentiated}. We put the factorization and the renormalization scales equal, $\mu_F=\mu_R$.
To compare our predictions with the forthcoming LHC data, we set the center-of-mass energy $\sqrt s$ at $14$ TeV and fix the cone size at the value $R=0.5$.
We use the PDF set MSTW2008nnlo~\cite{pdf} and the two-lop running
coupling with $\alpha_s(M_Z)=0.11707$. Following a recent CMS study~\cite{CMS}, we restrict the rapidities of
the Mueller-Navelet jets to the region $3 \leq |y_J | \leq 5$, with steps
$\Delta y_J$ equal to 0.5. We present the 
cross section ${\cal C}_0$, the azimuthal decorrelations ${\cal C}_1/{\cal C}_0$ and ${\cal C}_2/{\cal C}_0$ versus the relative rapidity.
\\
We perform our calculation both in the LLA and in the NLA. In the LLA we fixed the values of the renormalization and energy scales, $\mu_R$ and $s_0$, as suggested by the kinematics of the process, {\it i.e.} $\mu_R^2=s_0=|\vec k_{J_1}||\vec k_{J_2}|$. In the NLA, following Ref.~\cite{exponentiated},   we use an adaptation of the {\it principle of minimal sensitivity} (PMS)~\cite{PMS}, which consists in taking as optimal choices for $\mu_R$ and $s_0$ those values for which the physical observable  exhibits the minimal sensitivity to changes of both these scales.
The motivation of this procedure is that complete resummation of the perturbative series would not depend on the scales $\mu_R$ and $s_0$, so the optimization method is supposed to mimic the effect of the most relevant uncalculated  subleading terms. In our search for optimal values, we took integer values for $Y_0=\ln(s_0/(|\vec k_{J_1}| |\vec k_{J_2}|))$.
We found that, for ${\cal C}_0$, a stationary point in the $\mu_R$--$s_0$ plane could always be
singled out, typically a local maximum. As in previous works~\cite{exponentiated},
the optimal values of the energy scales turn to be far from the
kinematic scale.
\\
In a similar manner, we performed the analysis to determine   ${\cal C}_1/{\cal C}_0$ and ${\cal C}_2/{\cal C}_0$. In this case we found that the stability region is less evident than in ${\cal C}_0$.
\\
We studied Mueller-Navelet jets in the cases $|\vec k_{J_{1,2}}|=35$ GeV,  $|\vec k_{J_{1,2}}|=20$ GeV and $|\vec k_{J_1}|=20$ GeV and $|\vec k_{J_2}|=35$ GeV. For a summary of the results, see Figs. \ref{fig:C035}--\ref{fig:C02035}.
\vspace{-0.2cm}
\section{Conclusions}

We studied the $Y$-dependence of the cross section and the azimuthal decorrelation of the Mueller-Navelet jets in full NLA BFKL, at different values of jet transverse momenta, using the jet vertex in the small-cone approximation. An optimized procedure (PMS) has led to results stable in the considered energy interval. Our predictions at $|\vec k_{J_{1,2}}|=35$ GeV are compatible with previous calculation where the jet cone size was treated exactly~\cite{Colferai}. In particular we have found a complete agreement in ${\cal{C}}_0$ and a moderate discrepancy 
for the moments of the azimuthal decorrelations. So, our study substantially confirms the results of~\cite{Colferai} and demonstrates the reliability of the small-cone approximation.
Instead the analysis at $|\vec k_{J_{1,2}}|=20$ GeV is new.  In this kinematics more undetected gluon radiation is expected, which allows for a better discrimination between BFKL and NLO DGLAP approaches. 

\vspace{-0.2cm}
\begin{thebibliography}{9}

\bibitem{Mueller:1986ey}
A.H.~Mueller, H.~Navelet, Nucl. Phys. B {\bf 282} (1987)  727.

\bibitem{small_cone}
D.~Yu.~Ivanov, A.~Papa, JHEP {\bf 1205} (2012) 086.

\bibitem{BFKL}
V.S.~Fadin, E.A.~Kuraev, L.N.~Lipatov, Phys. Lett. {\bf B60} (1975) 50;
E.A.~Kuraev, L.N.~Lipatov and V.S.~Fadin, Zh. Eksp. Teor. Fiz. {\bf 71} (1976)
840 [Sov. Phys. JETP {\bf 44} (1976) 443]; {\bf 72} (1977) 377 [{\bf 45} (1977)
199];
Ya.Ya.~Balitskii and L.N.~Lipatov, Sov. J. Nucl. Phys. {\bf 28} (1978) 822.

\bibitem{Colferai}
D.~Colferai, F.~Schwennsen, L.~Szymanowski, S.~Wallon,
JHEP {\bf 1012 } (2010)  026.

\bibitem{DSW}
B.~Duclou\'e, L.~Szymanowski, S.~Wallon,  arXiv:1208.6111.

\bibitem{Sabio_Vera}
A.~Sabio Vera, Nucl. Phys. B {\bf 746} (2006) 1;
A.~Sabio Vera, F.~Schwennen, Nucl. Phys. B {\bf 776} (2007) 170;
C.~Marquet, C.~Royon, Phys. Rev. D {\bf 79} (2009) 034028.

\bibitem{exponentiated}
D.Yu.~Ivanov and A.~Papa, Nucl. Phys. B {\bf 732} (2006) 183;
Eur. Phys. J. C {\bf 49} (2007) 947; F.~Caporale, A.~Papa and A.~Sabio Vera,
Eur. Phys. J. C {\bf 53} (2008) 525.

\bibitem{pdf}
A.D. Martin, W.J. Stirling, R.S. Thorne and G. Watt, Eur. Phys. J. C {\bf 63}
(2009) 189.

\bibitem{CMS}
S. Cerci and D. d'Enterria,  AIP Conf. Proc. {\bf 1105} (2009) 28.

\bibitem{PMS}
P.M.~Stevenson, Phys. Lett. {\bf B100} (1981) 61;
Phys. Rev. D {\bf D23} (1981) 2916.

\end{thebibliography}
\end{document}